\documentclass[aps,prd,onecolumn]{revtex4}

\usepackage{graphicx}  % standard LaTeX graphics tool

 \usepackage{amssymb}
 
\begin{document}

\title{A possible approach to understand nonlinear gravitational clustering in expanding background }

\author{T.~Padmanabhan}
\affiliation{IUCAA,
Post Bag 4, Ganeshkhind, Pune - 411 007\\
email: nabhan@iucaa.ernet.in}

%%upright Greek letters (example below: upright "mu")
\newcommand{\greeksym}[1]{{\usefont{U}{psy}{m}{n}#1}}
\newcommand{\umu}{\mbox{\greeksym{m}}}
\newcommand{\udelta}{\mbox{\greeksym{d}}}
\newcommand{\uDelta}{\mbox{\greeksym{D}}}
\newcommand{\uPi}{\mbox{\greeksym{P}}}

%AUTHOR_STYLES_AND_DEFINITIONS%%%%%%%%%%%%%%%%%%%%%%%%%%%%%%%

  \def\bld#1{{\bf #1 }}
 \def\xb{\bar\xi(a,x)}
 \def\gaprox{\mbox{$\,$ 
\raisebox{0.5ex}{$<$}\hspace{-1.7ex}{\raisebox{-0.5ex}{$\sim$ }}$\,$} }
\def\part#1#2{\frac{\partial #1}{\partial #2}}
\def\pder#1#2{{\partial #1/\partial #2}}

\def\rb{\right)}
\def\lb{\left(}

\def\frab#1#2{\left({#1\over#2}\right)}
\def\fra#1#2{{#1\over#2}}

%%%%%%%%%%%%%%%%%%%%%%%%%%%%%%%%%%%%%%%%%%%%%%%%%%%%%%%%

\begin{abstract}
A new approach to study the nonlinear phase of gravitational clustering in an expanding universe is explored. This approach is based on 
an integro-differential equation for the evolution of the gravitational potential in the Fourier space which is obtained by using a physically motivated closure condition. I show how this equation allows one to understand several aspects of nonlinear gravitational clustering and provides insight in to the transfer of power from one scale to another through nonlinear mode coupling. 
\end{abstract}

\maketitle

  \section{\label{intro}Introduction }
  
 There is considerable amount of observational evidence to suggest that
 about 25 per cent of the energy density in the universe is contributed by self gravitating system of
  dark matter particles. The smooth, average, energy density of these particles contributes to
  the expansion of the universe while any small deviation from the homogeneous energy
  density will lead to gravitational clustering. One of the central problems in cosmology is to describe
  the non linear phases of this gravitational clustering starting from a initial spectrum of density
  fluctuations. It is often enough (and necessary) to use a statistical description and  relate
  different statistical indicators (like the  power spectra, $n$th order  correlation functions etc.)
  of the resulting density distribution to the statistical parameters (usually the power spectrum) of the 
  initial distribution. The relevant scales at which gravitational clustering is non linear are less than
  about 10 Mpc (where 1 Mpc $\approx 3\times 10^{24}$ cm is the typical separation between galaxies in the
  universe) while the expansion of the universe has a characteristic scale of about few thousand 
  Mpc. Hence, non linear gravitational clustering in an expanding universe can 
  be adequately described by Newtonian gravity provided the rescaling of lengths due to the
  background expansion
  is taken into account. 
  
 As to be expected, cosmological expansion  introduces  
  several new factors into the problem as compared to the study of statistical mechanics of isolated
  gravitating systems. (For a general review of statistical mechanics of gravitating systems,
  see \cite{tppr}. For a sample of different approaches, see \cite{sample} and the references cited therein. 
  Review of gravitational clustering in expanding background is  available in several textbooks in cosmology \cite{cosmotext,lssu}.)
  (a) The problem has now become time dependent and it will be pointless to look for
  equilibrium solutions in the conventional sense of the word. 
  (b) On the other hand, the expansion of the universe has a civilizing influence on the 
  particles and  acts counter to the tendency of gravity
  to make systems unstable. 
  (c) In any small local region of the universe, one would assume that the conclusions
  describing a finite gravitating system will still hold true approximately. In that case, particles in any small
  sub region will be driven towards configurations of local extrema of entropy (say, isothermal
  spheres) and towards global maxima of entropy (say, core-halo configurations). It is not clear how  these effects and cosmological expansion interact with each other at intermediate length scales.
  
 Though this problem can be
tackled in a `practical' manner using high resolution numerical simulations,
such an approach hides the physical principles which govern 
the behaviour of the system. To understand the physics, it is necessary to
attack the problem from several directions using analytic and semi analytic
methods. Several such attempts exist in the literature based on Zeldovich(like) approximations \cite{za},
path integral and perturbative techniques \cite{pi}, nonlinear scaling relations \cite{nsr} and many others. In spite of all these it is probably fair to say that we still do not have a clear analytic grasp of this problem, mainly because each of these approximations have different domains of validity and do not arise from a central paradigm.

The purpose of this paper is to attack the problem from a different angle, which has not received much attention in the past. The approach begins from the dynamical equation for the the density contrast in the Fourier space  and casts it as an integro-differential equation. This equation is known in the literature  (see, e.g. \cite{lssu}) but has received very little attention  because it is not `closed' mathematically; that is, it involves variables which are not natural to the formalism and thus further progress is difficult. I will, however, argue that
there exists a natural closure condition for this
equation  based on Zeldovich approximation thereby allowing us to write down
 a \textit{closed} integro-differential equation for the gravitational potential
in the Fourier space. It turns out that this equation can form the basis for several further investigations. In fact, one purpose of this paper
--- which is probably somewhat more pedagogical than is usual --- is to draw the attention of the community to this approach and encourage further investigations based on this equation and ansatz.

Given the validity of the ansatz, several conclusions follow in a straight forward manner. For example, I will describe
how different aspects of non linear gravitational clustering (which are well known in the 
literature) can all be obtained by taking suitable limits of this equation. What is more important, the structure of this integro-differential equation  allows one to understand the key issue of transfer of power from one scale to another in non linear
gravitational clustering. If the initial power spectrum is sharply peaked in Fourier space,
in a \textit{shell} of radius $|{\bf k}| = k_0$, then the non linear gravitational clustering allows for
both cascading (to smaller length scales) \textit{and inverse} cascading (to larger length scales) of the power during the evolution. The formalism developed here shows that
the inverse cascading leads to the well known $k^4$ tail at small wave numbers (that is, at large
spatial scales). Again, while this result is known, the derivation given here is new and appears to be simple, straightforward and does not require any other extra ad-hoc assumption.  The cascading of power, on the other hand, proceeds broadly through the generation of harmonics at
$2k_0, 4 k_0$ etc. with the mode coupling quickly leading to a universal power spectrum.
These results agree with numerical simulations done in the past \cite{jsbtp1}. Finally, it turns out that
one can obtain scale free solution to the integro-differential equation and analyse
its asymptotic properties completely. This study shows  some striking similarities between nonlinear gravitational clustering of collisionless particles in an expanding universe and standard fluid turbulence. This application as well as its verification in numerical 
simulations will be described in a separate paper \cite{stp}.

\section{\label{gravclnl}Nonlinear Gravitational clustering}

The expansion of the universe sets a natural length scale (called the Hubble radius) $d_H = c
(\dot a/a)^{-1}$ which is about 4000 Mpc in the current universe. Since the non linear effects due
to gravitational clustering occur at significantly smaller length scales, it is possible to use
Newtonian gravity to describe these phenomena. 
In any region which is small compared to $d_{\rm H}$ one can set up an unambiguous coordinate system in which the {\it proper} coordinate of a particle ${\bf r} (t)=a(t){\bf x}(t)$ satisfies the Newtonian equation $\ddot {\bf r} = -  {\nabla }_{\bf r}\Phi$ where $\Phi$ is the gravitational potential. Expanding $\ddot \bld r$ and writing $\Phi = \Phi_{\rm FRW} + \phi$ where $\Phi_{\rm FRW}$ is due to the smooth (mean) density 
of matter  and $\phi$ is due to the perturbation in the density, we get
\begin{equation}
\ddot a {\bf x} + 2 \dot a \dot{\bf x} + a\ddot{\bf x} = - \nabla_{\bf r} \Phi_{\rm FRW} - \nabla_{\bf r}\phi = - \nabla_{\bf r} \Phi_{\rm FRW} - a^{-1} \nabla_{\bf x} \phi
\end{equation}
The first terms on both sides of the equation $\lb \ddot a{\bf x} \  {\rm and} -\nabla_{\bld r} \Phi_{\rm FRW} \rb$ should match since they refer to the global expansion of the background FRW universe. Equating them individually gives the results
\begin{equation} 
\ddot{\bf x} + 2 {\dot a \over a}\dot{\bf x} = - {1 \over a^2} \nabla_x \phi\ ; \qquad \Phi_{\rm FRW} = - {1 \over 2}{\ddot a \over a} r^2 = - {2\pi G \over 3}\rho_b r^2 
\end{equation}
where $\phi$ is the gravitational potential generated by the \textit{perturbed}  mass density $\delta \lb t, \bld x \rb \equiv
[\rho(t,{\bf x}) - \rho_b(t)]/\rho_b(t)$ through 
\begin{equation}
 \nabla^2_x \phi = 4 \pi Ga^2(\delta \rho) = 4 \pi G \rho_ba^2 \delta 
  \end{equation}
(The $\Phi_{FRW}$ is due to a uniform density sphere, as to be expected.)
Hence, the equations for 
 gravitational clustering in an expanding universe, in the Newtonian
 limit, can be summarized by
\begin{equation}
\ddot{\bf x}_i + { 2\dot a \over a} \dot{\bf x}_i = - {1 \over a^2} \nabla_{\bf x}
\phi;\quad \nabla_x^2 \phi = 4\pi G a^2 \rho_b \delta  \label{twnine}
\end{equation}
where $\rho_b(t)$ is the smooth background density of matter and ${\bf x}_i(t)$ is the trajectory of the $i-$th particle. We stress that, in the non-relativistic limit,
 the perturbed potential $\phi$ satisfies the usual Poisson equation with the perturbed density contrast as the source.

Usually one is interested in the evolution of the density contrast $\delta(t,{\bf x})$ rather than in the trajectories. Since the density contrast can be expressed in terms of the trajectories of the particles, it should be possible to write down a differential equation for $\delta (t, \bld x)$ based on the equations for the trajectories $\bld x_i (t)$ derived above. It is, however, somewhat easier to write down an equation for $\delta_{\bld k} (t)$ which is the spatial Fourier transform of $\delta (t, \bld x)$. To do this, we begin with the fact that the density $\rho(\bld x,t)$ due to a set of point particles, each of mass $m$, is given by
\begin{equation}
\rho (\bld x,t) = {m\over a^3 (t)} \sum\limits_i \delta_D [ \bld x - \bld x _i(t)]
\end{equation}
where $\bld x_{i}(t)$ is the trajectory of the ith particle. To verify the $a^{-3}$ normalization, we can calculate the average of $\rho(\bld x,t)$ over a large volume $V$. We get
\begin{equation}
\rho_b(t) \equiv \int  {d^3 \bld x \over V} \rho (\bld x, t) = {m\over a^3(t)} \lb {N\over V}\rb = {M\over a^3 V} = {\rho_0\over a^3}
\end{equation}
where $N$ is the total number of particles inside the volume $V$ and $M = Nm$ is the mass contributed by them. Clearly $\rho_b \propto a^{-3}$, as it should.  The density contrast $\delta (\bld x,t)$ is related to $\rho(\bld x, t)$ by
\begin{equation}
1+\delta (\bld x,t) \equiv {\rho(\bld x, t) \over \rho_b} = {V \over N} \sum\limits_i \delta_D [\bld x - \bld x_i(t)] =  \int d {\bld q} \delta_D [\bld x - \bld x_{T} (t, \bld q)]  . 
\end{equation}
In arriving at the last equality we have taken the continuum limit by two steps: (i) We have replaced $\bld x_i(t)$ by $\bld x_T(t,\bld q)$ where  $\bld q$ stands for a set of parameters (like the initial position, velocity etc.) of a particle; for
simplicity, we shall take  this to be initial position. The subscript `T' is just to remind ourselves that $\bld x_T(t,\bld q)$ is the
\textit{trajectory} of the particle. (ii) We have also replaced $(V/N)$ by $d^3{\bld q}$ since both represent volume per particle. Fourier transforming both sides  we get
\begin{equation}
\delta_{\bld k}(t) \equiv \int d^3\bld x   {\rm e}^{-i\bld k \cdot \bld x} \delta (\bld x,t) =   \int d^3 {\bld q} \  {\rm exp}[ - i {\bf k} . {\bf x}_{T} (t, \bld q)]  -(2 \pi)^3 \delta_D (\bld k)
\end{equation}
Differentiating this expression, 
and using  Eq.~(\ref{twnine}) for the trajectories give, after straightforward algebra, the equation (see  \cite{pines},  \cite{tpprob},\cite{lssu}): 
\begin{equation}
\ddot \delta_{\bf k} + 2 {\dot a \over a} \dot \delta_{\bf k} = {1\over a^2}\int d^3 \bld q 
e^{-i{\bf k}. {\bf x }_T(t, \bld q) }\left\{ i{\bf k}\cdot \nabla \phi - a^2 ({\bf k \cdot \dot  x}_T)^2\right\}
 \label{basic}
\end{equation}
which can be further manipulated to give
\begin{equation}
\ddot \delta_{\bf k} + 2 {\dot a \over a} \dot \delta_{\bf k} = 4 \pi G \rho_b \delta_{\bf k} + A _{\bld k}- B_{\bld k} \label{exev}
\end{equation}
with
\begin{equation} 
A_{\bld k} =4\pi  G\rho_b \int{d^3{\bf k}' \over (2 \pi)^3}  \delta_{\bf k'} \delta_{{\bf k} - {\bf k'}} \left[{{\bf k}. {\bf k'} \over k^{'2}} \right] 
\end{equation}
\begin{equation}
B_{\bld k} = \int d^3 \bld q    \left({\bf k}.{\dot{\bf x}_T}  \right)^2 {\rm exp} \left[ -i{\bf k}. {\bf x }_T(t, \bld q) \right] .\label{exevii} 
\end{equation}
This equation is exact but involves ${\bf x}_T (t,{\bf q}) $ and$\dot{\bf x}_{T}(t, \bld q)$  on the right hand side;   hence it  cannot be considered as closed. 
 
The structure of Eq.~(\ref{exev})  can be simplified if we use the perturbed gravitational potential (in Fourier space) $\phi_{\bf k}$ related to $\delta_{\bf k}$ by 
\begin{equation}
\delta_{\bf k} = - {k^2\phi_{\bld k} \over 4 \pi G \rho_b a^2} = - \lb {k^2 a \over 4 \pi G \rho_0}\rb \phi_{\bld k} = - \lb {2 \over 3H_0^2 }\rb k^2a \phi_{\bld k}
\end{equation}
and write the integrand for $A_{\bld k}$ in the symmetrised form as 
\begin{eqnarray}
\delta_{\bld k'} \delta_{\bld k - \bld k'} \left[ {\bld k . \bld k' \over k^{'2}} \right]& = &{1 \over 2} \delta_{\bld k'} \delta_{\bld k - \bld k'}\left[  {\bld k . \bld k' \over k^{'2}}  + {\bld k . (\bld k - \bld k') \over | \bld k - \bld k'|^2} \right] \nonumber \\
&=& { 1\over 2} \left( {\delta_{\bld k}'} \over k^{'2} \right) \left( {\delta_{\bld k - \bld k'} \over | \bld k - \bld k'|^2} \right) \left[ (\bld k - \bld k')^2 \bld k . \bld k' + k^{'2}\left( k^2 - \bld k . \bld k'\right)\right]\nonumber \\
&=& {1\over 2} \left({2a \over 3H_0^2}\right)^2 \phi_{\bld k'} \phi_{\bld k - \bld k'} \left[ k^2 (\bld k . \bld k' + k^{'2}) - 2(\bld k . \bld k')^2 \right] \nonumber \\
\end{eqnarray}
In terms of $\phi_{\bld k}$ (with ${\bf k}' = ({\bf k}/2) +{\bf p})$, equation (\ref{exev}) becomes, 
\begin{eqnarray}
\ddot \phi_{\bf k} + 4 {\dot a \over a} \dot\phi_{\bf k}   &= & - {1 \over 2a^2} \int {d^3{\bf p} \over (2 \pi )^3} \phi_{{1\over 2}{\bf k+p}}   
\phi_{{1\over 2}{\bf k-p}}\left[ \left( {k\over 2}\right)^2 + p^2 
-2  \lb {\bld k . \bld p \over k}\rb^2 \right] \nonumber \\
&+ &\lb{3H_0^2 \over 2}\rb  \int {d^3{\bf q} \over a} \lb{\bld k} . \dot {\bld x}\over k\rb ^2 e^{i{\bf k}.{\bf x}} \label{powtransf} 
\end{eqnarray}
where $\bld x = \bld x_T(t, \bld q)$. 

 Of course, this equation is not `closed' either.
 It contains the velocities of the particles $\dot {\bf x}_T$ and their positions explicitly in the second term on the right and one cannot --- in general --- express them in simple form in terms of $\phi_\mathbf{k}$. As a result, it might seem that we are in no better position than when we started. I will now motivate a strategy to tame this term in order to close this equation. This strategy  depends on two features: First, extremely nonlinear structures do not contribute to the difference $(A_{\bf k}-B_{\bf k})$ though, of course, they contribute individually to both $A_{\bf k}$ and $B_{\bf k}$. Second, we can use Zeldovich approximation to evaluate this term, once the above fact is realised. I will now elaborate on these two features.

  \subsection{\label{renormgrav} `Renormalizability' of gravity}
  
  Gravitational clustering in an expanding universe brings out an interesting feature 
  about gravity which can be described along the following lines. Let us consider a large
  number of particles which are  interacting via gravity in an expanding background and forming
  bound gravitating systems. At some time $t$, let us assume that a fraction $f$ of the particles are
  in virialized, self-gravitating clusters (of typical size $R$, say) which are reasonably immune to the effect of 
  expansion. Imagine that we replace each cluster by a single particle
  at its  center of mass with the mass equal to the total mass of the  cluster. 
  (The total number of 
  particles have now been reduced but, if the original number was sufficiently large, we may
  assume that the resulting number of particles is again large enough to carry on further evolution
  with a valid statistical description.) We now evolve the resulting system to a time $t'$ and 
  compare the result with what would have been obtained if we had evolved the original system
  directly to $t'$. Obviously, the characteristics of the system at small scales (corresponding to the
  typical size $R$ of the clusters at time $t$) will be quite different.
  However, at large scales ($kR\ll 1$), the characteristics will be the same both the systems. In other words,
  the effect of a bunch of particles, in a virialized cluster, on the rest of the system
  is described, to the lowest order, by just the monopole moment of the cluster --
  which is taken into account by replacing the cluster by a single particle at the center of mass
  having appropriate mass. In this sense, gravitational interactions are ``renormalizable'' --
  where the term is used in the specific sense defined above.
  
  The result has been explicitly verified in simulations  but one must emphasize that
  the whole idea of numerical simulations of such systems tacitly \textit{assumes} the validity of this 
  result. If the detailed non linear behaviour at small scales, say within galaxies, influences
  very large scale behaviour of the universe (say, at super cluster scales), then it will be
  impossible to study the large scale structure in the universe with simulations of finite resolution.
 
  One may wonder how this feature (renormalizability of gravity) is taken care
  of in Eq.~(\ref{exev}). Inside a galaxy cluster, for example, the velocities
  ${\bf \dot x}_T$  can be quite high and one might think that this could influence
  the evolution of $\delta_{\bf k}$ at all scales. This does not happen and, to the lowest order,
  the contribution from virialized bound clusters cancel in $A_{\bf k} - B_{\bf k}$.  
  We shall now provide a proof of this result (also see \cite {lssu}). 
  
  We begin by  writing the right hand side ${\cal R}$ of the 
   Eq.~(\ref{basic}) concentrating on the particles in a given cluster.
  \begin{equation}
 {\cal R}  = \int d^3 {\bf q}\,  e^{-i{\bf k\cdot Q}} i k^a \left\{ 
  {\partial_a \phi\over a^2} + i \dot Q^a ({\bf k \cdot \dot Q})\right\}
  \label{altone}
  \end{equation} 
  where we have used the notation ${\bf Q} = {\bf x}_T$ for the trajectories of the particles
  and the subscripts $a, b, .... = 1, 2, 3 $ denote the components of the vector.
  For a set of particles which form a bound virialized cluster, we have from Eq.~(\ref{twnine}) the
  equation of motion 
  \begin{equation}
  \ddot Q^i + 2 {\dot a\over a} \dot Q^i = - {1\over a^2} {\partial \phi\over \partial Q^i}
  \end{equation}
  We multiply this equation by
  $Q^j $, sum over the particles in the particular cluster and symmetrize on $i $ and $j$, to obtain
  the equation 
  \begin{equation}
  {d^2\over dt^2} \sum Q^i Q^j - 2 \sum \dot Q^i \dot Q^j + 2 {\dot a\over a} {d\over dt}
  \sum Q^iQ^j = -{1\over a^2} \sum \left( Q^j {\partial \phi \over \partial Q^i} + Q^i {\partial \phi\over 
  \partial Q^j}\right)
  \label{alttwo}
  \end{equation}
 We use the summation symbol, rather than integration over ${\bf q}$ merely to emphasize the fact that the sum is over particles of a \textit{given} cluster. Let us now consider the first term in the right hand side of  Eq.~(\ref{altone}) with the origin
  of the coordinate system shifted to the center of mass of the cluster. Expanding the exponential
  as $e^{-i{\bf k\cdot Q}} \approx (1-i{\bf k\cdot Q}) + {\cal O}(k^2 R^2) $ where $R$
  is the size of the cluster, we find that in the first term,  proportional to $\nabla \phi$, 
    the sum of the forces acting on
  all the particles in the cluster (due to self gravity) vanishes. The second term gives, on symmetrization,
  \begin{equation}
  \sum i a^{-2}({\bf k\cdot }\nabla \phi)e^{-i{\bf k\cdot Q}} \approx {k^ak^b\over 2a^2} \sum \left( Q^b 
  {\partial \phi\over \partial Q^a} + Q^a  {\partial \phi\over \partial Q^b}\right)
  \end{equation}
  Using Eq.~(\ref{alttwo}) we find that 
  \begin{equation}
   \sum i a^{-2}({\bf k\cdot }\nabla \phi)e^{-i{\bf k\cdot Q}} = + \sum ({\bf k\cdot \dot Q})^2 + {1\over 2}
   \left( {d^2 \over dt^2} + 2 {\dot a\over a} {d\over dt}\right) \sum ({\bf k\cdot Q})^2
   \end{equation}
   The second term is of order ${\cal O}(k^2R^2)$ and can be ignored, giving
   \begin{equation}
   \sum i a^{-2}({\bf k\cdot }\nabla \phi)e^{-i{\bf k\cdot Q}} \approx + \sum ({\bf k\cdot \dot Q})^2
    +{\cal O}(k^2R^2)
   \end{equation}
   Consider next the second term in the right hand side of (\ref{altone}) with the same
   expansion for the exponential. We get
   \begin{eqnarray}
   \sum (ik^a) e^{-i{\bf k\cdot Q}} &\left[ i \dot Q^a k^b \dot Q^b\right] \cong - \sum k^ak^b \dot Q^a
   \dot Q^b ( 1- i {\bf k\cdot Q}) +{\cal O}(k^2R^2)\nonumber \\
   &= - \sum ({\bf k\cdot \dot Q})^2 + \sum ({\bf k\cdot \dot Q})^2 k^a Q^a +{\cal O}(k^2R^2)
   \end{eqnarray}
   The second term is effectively zero for any cluster of particles for which ${\bf Q}\to -{\bf Q}$
   is a symmetry. Hence the two terms on the right hand side of Eq.~(\ref{altone}) cancel each other
   for all particles in the same virialized cluster; that is,  to the order
   ${\cal O}(k^2R^2)$, the term $(A_{\bf k} - B_{\bf k})$ receives
   contribution only from particles which are not bound to any of the clusters. If the typical size of the clusters formed at time $t$
   is $R$, then for wave-numbers with $k^2 R^2 \ll 1$, we can ignore the contribution from the clusters.
   Hence, in the limit of $k \to 0$ we can ignore $(A_{\bf k} -B_{\bf k})$
   term and treat equation Eq.~(\ref{exev}) as linear in $\delta_{\bf k}$; 
    large spatial scales in the universe can be described by linear perturbation theory
   even when small spatial scales are highly non linear.
   
   There is, however, an important caveat to this claim. In the right hand side of Eq.~(\ref{exev}) one 
   is comparing the first term (which is linear in $\delta_{\bf k}$) with  the contribution
   $(A_{\bf k} -B_{\bf k})$. If, at the relevant wavenumber, the first term  $4\pi G \rho_b \delta_{\bf k}$
   is negligibly small, then
   the {\it only} contribution will come from $(A_{\bf k} -B_{\bf k})$ and, of course, we cannot
   ignore it in this case. The above discussion shows that this contribution will scale as 
   $k^2 R^2$ and will lead to a development of $\delta_{\bf k} \propto k^2$ if originally (in linear 
   theory) $\delta_k \propto k^n$ with $n>2$ as $k\to 0$.  We shall say more about this later on.

Clearly, we can ignore the contribution from particles in virialized clusters in estimating $(A_\mathbf{k}-B_\mathbf{k})$. We will next consider how we can estimate the effect of remaining particles.
A useful insight  can be
obtained by examining the nature of particle trajectories which lead
to the growth of the density contrast $\delta_{\bf k}\propto a$  in the linear limit.
To determine the particle trajectories corresponding to the 
linear limit, let us start by writing the trajectories in the form
\begin{equation} 
{\bf x}_T (a,{\bf q}) = {\bf q} + {\bf L} (a,{\bf q})
\end{equation}
where ${\bf q}$ is the Lagrangian coordinate (indicating the 
original position of the particle) and ${\bf L}(a,{\bf q})$ is 
the displacement. The corresponding Fourier transform of the density contrast is given by the general expression
\begin{equation}
 \delta_{\bf k} (a)= \int d^3{\bf q}\, e^{-i{\bf k\cdot q}-i{\bf k\cdot L}(a,{\bf q})} - (2 \pi)^3 \delta_{\rm Dirac} [{\bf k}]
\end{equation}
In the linear regime, we expect the particles to have moved very little
and hence we can expand the integrand in the above equation in a Taylor
series in $({\bf k\cdot L})$. This gives, to the lowest order, 
\begin{equation} 
\delta_{\bf k} (a)\cong -\int d^3{\bf q}\, e^{-i{\bf k\cdot q}} (i{\bf k\cdot L}(a,{\bf q})) = -\int d^3{\bf q}\, e^{-i{\bf k\cdot q}}\lb \nabla_{\bf q} \cdot {\bf L}\rb
\end{equation}
showing that $\delta_{\bf k}(a)$ is Fourier transform of $-\nabla_{\bld q} . \bld L (a, \bld q)$. This allows  us to identify $\nabla\cdot {\bf L}(a,{\bf q})$ with
the original density contrast in real space $- \delta_{\bf q} (a)$. Using
the Poisson equation  we can write $\delta_{\bf q}(a)$ as a divergence; that is 
\begin{equation} 
\nabla \cdot {\bf L}(a,{\bf q}) = - \delta_{\bf q}(a) = - \fra{2}{3} H_0^{-2} a \nabla \cdot (\nabla \phi)
\end{equation}
which, in turn, shows that    a consistent set of displacements that will
lead to $\delta (a) \propto a$ is given by
\begin{equation}
{\bf L}(a,{\bf q}) = -  (\nabla \psi)a \equiv a {\bf u}({\bf q}) ; 
   \qquad \psi\equiv (2/3) H_0^{-2}\phi \label{ninety}
\end{equation} 
The trajectories in this limit
are, therefore, linear in $a$: 
\begin{equation}
 \bld x_{T} (a,{\bf q}) = {\bf q} + a {\bf u}({\bf q})\label{trajec}
\end{equation} 

A useful approximation to describe the quasi linear stages of clustering is obtained by using the trajectory in Eq.(\ref{trajec})  as an ansatz valid {\it even at quasi linear epochs}. In this approximation, (called  Zeldovich approximation), the proper Eulerian position $\bld r $ of a particle is related to its Lagrangian position $\bld q $ by 
\begin{equation}
{\bf r}(t) \equiv a(t) {\bf x}(t) = a(t) [{\bf q} +  
a(t) {\bf u}({\bf q}) ] \label{lagq}  
\end{equation}
where ${\bf x}(t)$ is the comoving Eulerian coordinate.
 If the initial, unperturbed, 
density is $\overline \rho$ (which is independent of ${\bf q})$,
then the conservation of mass implies that the perturbed density will be
$
\rho ({\bf r},t) d^3{\bf r} = \bar \rho d^3{
\bf q}.\label{qmcons}
$
As is well known, this suggests that 
sheet like structures, or `pancakes', will
be the first nonlinear structures to form when gravitational instability
amplifies density perturbations.

 \subsection{Closure ansatz for the dynamical equation}
  
We now combine the two results obtained above in order to suggest a closure condition
for our dynamical equation. We begin by noting that at any given moment of time
we can divide the particles in the system into two  sets - those which are 
already a part of virialized cluster and those which are not. Of these, we know that the first  set  of particles do not
contribute significantly to $(A_{\bf k} - B_{\bf k})$ so we will not incur any serious error in ignoring these particles in computing
$(A_{\bf k} - B_{\bf k})$. For the description of particles 
in the second set, the Zeldovich approximation should be fairly good. In fact, we can do slightly
better than the standard Zeldovich approximation. We note that in Eq.~(\ref{lagq}) the velocities were
taken to be proportional to the gradient of the \textit{initial} gravitational potential.
We can improve on this ansatz by taking the velocities to be given by the gradient of the 
\textit{instantaneous} gravitational potential which has the effect of incorporating the
influence of particles in bound clusters on the rest of the particles to certain extent.
Given this ansatz, it is straightforward to obtain a \textit{closed} integro-differential
equation for the gravitational potential along the following lines.
 The trajectories in Zeldovich approximation, given by  Eq.~(\ref{trajec}) leads to:
\begin{equation}
\bld x_T(\bld q, a) = \bld q + a \nabla \psi ; \quad \dot \bld x_{\rm T} = \lb {2a \over 3t}\rb \nabla \psi; \quad \psi = {2 \over 3H_0^2 } \varphi
\end{equation}
To the same order of accuracy, $B_{\bld k}$ in Eq.~(\ref{exevii}) becomes:
\begin{equation}
\int d^3 \bld q \lb \bld k \cdot \dot\bld x_{\rm T}\rb^2e^{-i \bld k \cdot(\bld q + \bld L)} \cong \int d^3 \bld q ( \bld k \cdot \dot \bld x_{\rm T})^2 e^{-i \bld k \cdot \bld q}
\end{equation}
Substituting these expressions in Eq.~(\ref{powtransf}) we find that the gravitational potential is described by the closed integral equation:
\begin{eqnarray}
\ddot \phi_{\bld k} + 4 {\dot a \over a} \dot \phi_{\bld k} &=& -{1 \over 3a^2} \int {d^3 \bld p \over (2 \pi)^3} \phi_{{1 \over 2} \bld k + \bld p} \phi_{{1 \over 2} \bld k - \bld p} {\cal G} (\bld k, \bld p)\nonumber \\
{\cal G} (\bld k, \bld p) &= &{7 \over 8} k^2 + {3 \over 2} p^2 - 5 \lb {\bld k \cdot \bld p\over k}\rb^2 \label{calgxx} \nonumber \\
\end{eqnarray}
This equation provides a powerful method for analyzing non linear clustering since estimating $(A_{\bld k}-B_{\bld k})$ by Zeldovich approximation has a very large domain of applicability.

At this stage, the validity of the above equation rests on the conjectured validity of the 
ansatz used in approximating the trajectories. Though I have given a brief argument to motivate
this ansatz, its ultimate validity can be tested only by numerical simulations as well as by
working out the consequences of the equations. The rest of the paper works out several analytical
results which can be obtained from the above equation. We have also performed detailed numerical
simulations to check the above ansatz and some of these results will be presented elsewhere \cite{stp}.
For the propose of this paper, I merely state that the numerical simulations show that
this ansatz has a wide domain of applicability.

In the next two sections, I will use this equation to study the following important questions:
Suppose power was injected into a self gravitating system at a fixed length scale [that is, the 
initial power spectrum is $P_{\rm in}({\bf k}) \propto \delta_D(|{\bf k}| -k_0)$]. How does
the non linear evolution transfer the power to other wave numbers? The corresponding question in
fluid turbulence is very well studied and we know that for a wide range of scales the final spectrum
is the standard Kolmogorov spectrum. In the next two sections, we shall investigate some aspects
of this problem using the above formalism.

 \section{\label{nltail}
 Inverse cascade in non linear gravitational clustering: The $k^4$ tail}

  There is an interesting and curious result which is characteristic of gravitational
  clustering that can be obtained directly from our Eq.~(\ref{powtransf}). Consider an initial
  power spectrum which has very little power at large scales; more precisely, we shall
  assume that $P(k)$ dies faster than $k^4$ for small $k$. If these large scales are described
  by linear theory --- as one would have normally expected --- then the power at these scales 
  can only grow as $a^2$ and it will always be sub dominant to $k^4$. It turns out that this 
  conclusion is incorrect.   
As the system evolves, small scale nonlinearities will
develop in the system and --- if the large scales have too little
power intrinsically (i.e. if $n$ is large) ---  then
the long wavelength power will soon be dominated by the
``tail'' of the short wavelength power arising from the
nonlinear clustering. This occurs because, in Eq.~(\ref{exev}), the nonlinear term 
 ($A_{\bld k}$ - $B_{\bld k} )= {\cal O}(k^2R^2)$ can dominate over $4 \pi G \rho_b\delta_{\bld k}$ at long wavelengths  (as ${\bld k} \rightarrow 0$) and   lead to the development of a $k^4$ power spectrum at
  large scales. This is a purely non linear effect which we shall now describe.
  
A  formal way of obtaining the $k^4$ tail is to solve  Eq.~(\ref{calgxx}) for long wavelengths; i.e. near $\bld k = 0$.   Writing $\phi_{\bld k} = \phi_{\bld k}^{(1)} + \phi_{\bld k}^{(2)} + ....$ where $\phi_{\bld k}^{(1)} = \phi_{\bld k}^{(L)}$ is the time {\it independent} gravitational potential in the linear theory and $\phi_{\bld k}^{(2)}$ is the next order correction, we get from Eq.~(\ref{calgxx}), the equation  
\begin{equation}
\ddot\phi_{\bld k}^{(2)}+ 4 {\dot a \over a} \dot\phi_{\bld k}^{(2)} \cong - { 1 \over 3a^2} \int {d^3 \bld p \over (2 \pi)^3} \phi^L_{{1 \over 2} \bld k + \bld p} 
\phi^L_{{1 \over 2} \bld k - \bld p} {\cal G}(\bld k, \bld p)
\end{equation}
The solution to this equation is the sum of a solution to the homogeneous part [which decays as 
$\dot\phi\propto a^{-4}\propto t^{-8/3}$ giving $\phi\propto t^{-5/3}$] and a particular solution which grows as $a$. Ignoring the decaying mode at late times and taking
$\phi_{\bld k}^{(2)} = aC_{\bld k}$ one can determine $C_{\bld k}$ from the above equation. Plugging it back, we find the lowest order correction to be,
\begin{equation}
\phi_{\bld k}^{(2)} \cong - \lb {2a \over 21H^2_0}\rb \int {d^3 \bld p \over (2 \pi)^3}\phi^L_{{1 \over 2} \bld k + \bld p} 
\phi^L_{{1 \over 2} \bld k - \bld p} {\cal G}(\bld k, \bld p)
\label{approsol}
\end{equation}
Near $\bld k \simeq 0$, we have
\begin{eqnarray}
\phi_{\bld k \simeq 0}^{(2)} &\cong& - {2a \over 21H^2_0} \int {d^3 \bld p \over (2 \pi)^3}|\phi^L_{\bld p}|^2 \left[ {7 \over 8}k^2 + {3 \over 2}p^2 - {5(\bld k \cdot \bld p)^2 \over k^2} \right] \nonumber \\
&=&   {a \over 126 \pi^2H_0^2} \int\limits^{\infty}_0 dp  p^4 |\phi^{(L)}_{\bld p}|^2\nonumber \\
\end{eqnarray}
which is independent of $\bld k$ to the lowest order. Correspondingly the power spectrum for
 density $P_{\delta}(k)\propto a^2 k^4P_{\varphi} (k) \propto a^4 k^4$ in this order. 

\begin{figure}[ht]
\begin{center}
\includegraphics[width=.8\textwidth]{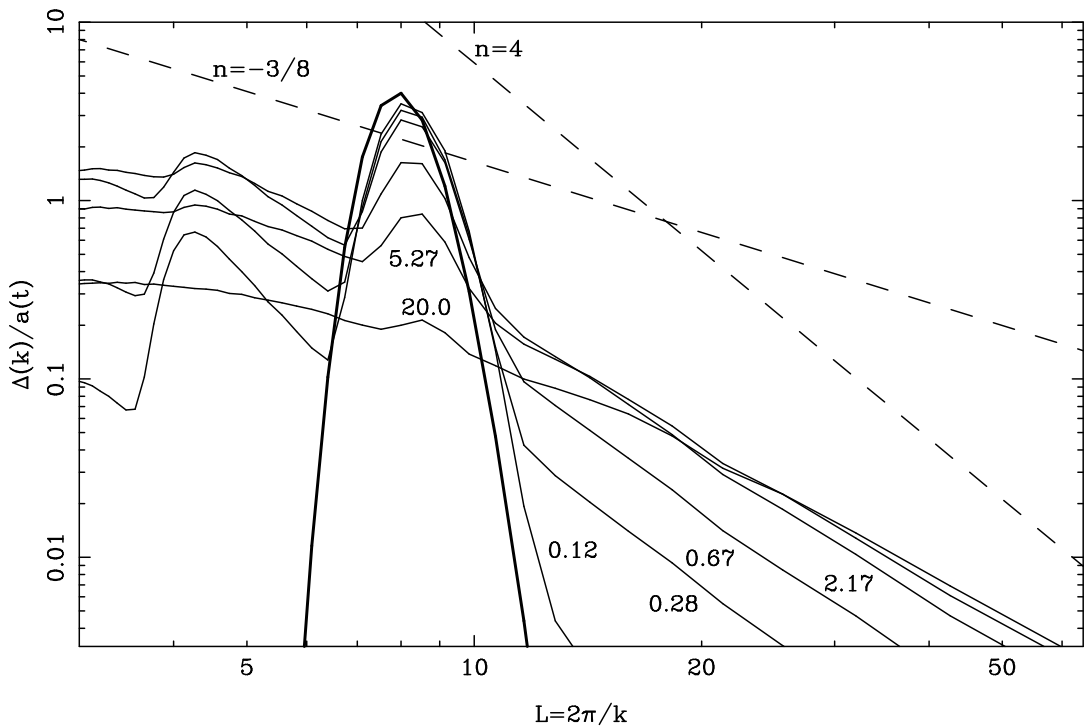}
\end{center}
\caption{The transfer of power to long wavelengths forming a $k^4$ tail is illustrated using 
simulation results. Power is injected in the form of a narrow peak at $L=8$. 
Note that the $y-$axis is $(\Delta/a)$ so that
there will be no change of shape of the power spectrum under linear evolution
with $\Delta\propto a$. As time goes on a $k^4$ tail is generated purely due to nonlinear coupling between the modes. (Figure adapted from ref.\cite{jsbtp1}.)}
\label{figptsimu}
\end{figure}

The generation of long wavelength $k^4$ tail is easily seen in simulations if one starts with a power spectrum that is sharply peaked in $|\bld k|$. Figure \ref{figptsimu}  shows the results of such a simulation (adapted from \cite{jsbtp1})  in which the y-axis is $[\Delta(k)/a(t)]$
where $\Delta^2(k) \equiv k^3P/2\pi^2$ is the power per logarithmic band in $k$. 
 In linear theory $\Delta \propto a$ and this quantity should not change. The curves labelled by $a=0.12$ to $a=20.0$ show the \textit{effects of nonlinear evolution}, especially the development of $k^4$ tail.

%%%%%%%%%% Figure  %%%%%%%%%%%%%%%%%%%%%%
%%%%%%%%%% End   Figure  %%%%%%%%%%%%%%%%%%%%%%

   %%%%%%%%%%%%%%%%%%%%%%%%%%%%%%%%%%%%% 
     \begin{figure}[ht]
   \begin{center}
   \includegraphics[width=0.85\textwidth]{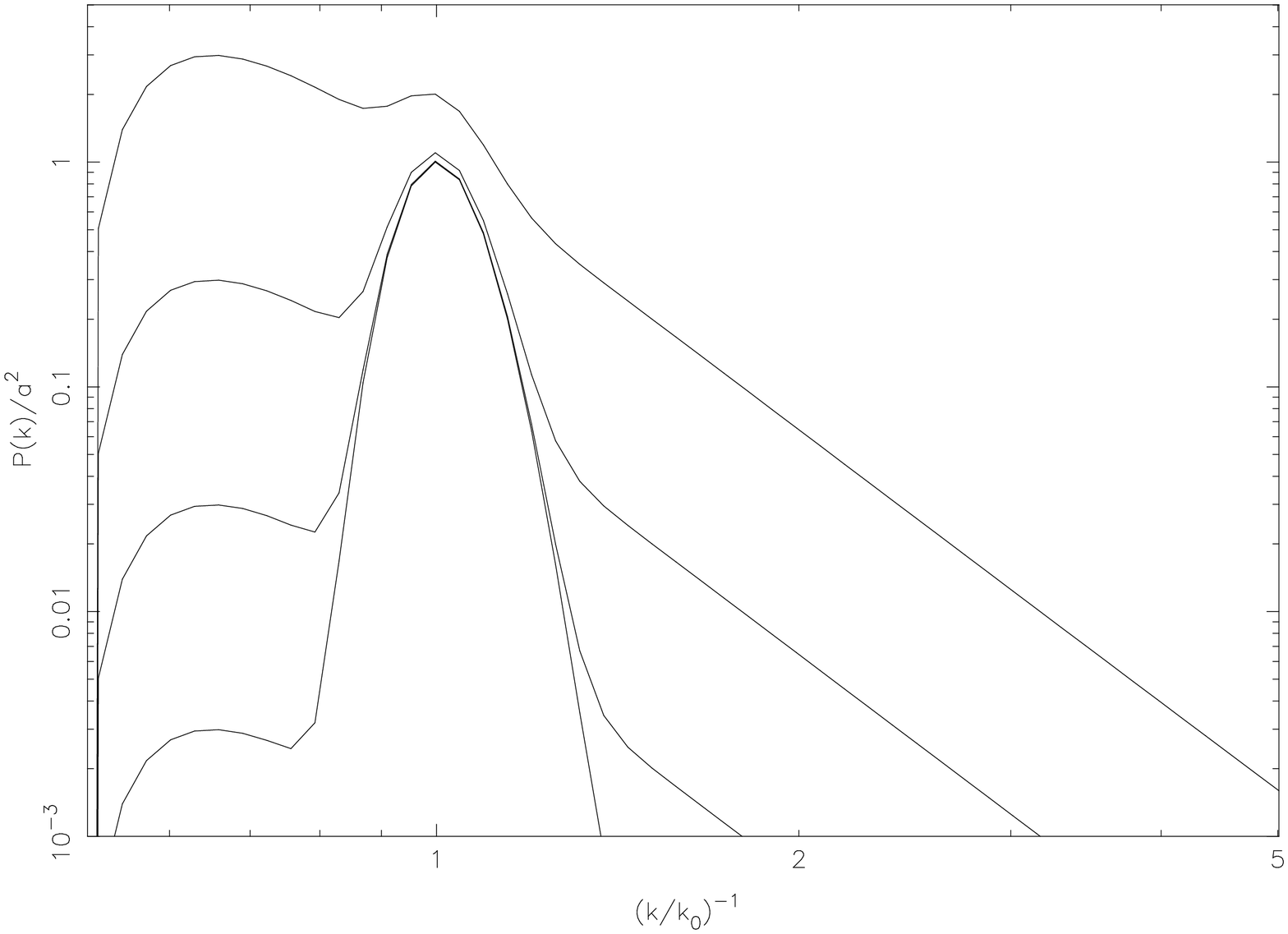}
   \end{center}
   \caption{Analytic model for transfer of power in gravitational clustering. The initial power
    was injected at the wave number $k_0$ with a Gaussian window of width $\Delta k/k_0= 0.1$.
    First order calculation shows that the power is transfered to larger spatial scales with a 
    $k^4$ tail and to the shorter spatial scales, all the way down to  $(1/2)k_0^{-1}$. The plot gives the
    total power spectrum divided by $a^2$ (with $y-$axis normalized arbitrarily) at different times with $a^2$
    changing by factor 10  between any two curves.}
    \label{figurepttheory}
    \end{figure}
    
    %%%%%%%%%%%%%%%%%%%%%%%%%%%%%%%%%%

\section{\label{gensmall}Cascading in non linear gravitational clustering: Generation of small scale power}

We next turn to the transfer of power to smaller spatial scales due to nonlinear mode coupling.
Figure \ref{figptsimu}   also shows that, as the clustering proceeds, power is generated at  spatial
scales smaller than the scale $k_0^{-1}$ at which the power is injected. (One can also see a characteristic peak 
at smaller scales.)
These features can 
also be easily understood  from   Eq.~(\ref{approsol}). Let the initial gravitational potential and
the density contrast 
(in the linear theory)  be sharply peaked at the wave number $k_0$, say, with:
\begin{equation}
\phi^L_{\bf k} = \mu {H_0^2\over k_0^4} \delta_D [|{\bf k}| - k_0]; \quad k_0^3 \delta^L_{\bf k} =-{2\over 3}
(\mu a) k_0 \delta_D [|{\bf k}| - k_0] 
\label{initcond}
\end{equation}
where $\mu$ is dimensionless constant indicating the strength of the potential and the other factors ensure the correct dimensions. Equation~(\ref{approsol}) shows that, the right hand side is nonzero only when the
{\it magnitudes} of both the vectors $[(1/ 2){\bf k}+{\bf p}]$ and $[(1/ 2){\bf k}-{\bf p}]$ are $k_0$. This
requires ${\bf k}\cdot{\bf p}=0$ and $ (k/2)^2+p^2=k_0^2$. 
(Incidentally, this constraint  has a simple geometric interpretation: Given any
${\bf k}$, with $k\le 2k_0$ one constructs a vector ${\bf k}/2$ inside a sphere of radius $k_0$ and
a vector ${\bf p}$ perpendicular to ${\bf k}/2$  reaching up to the shell at radius $k_0$
where the initial power resides. Obviously, this construction is possible only for
   $k< 2k_0$.) Performing the integration in Eq.~(\ref{approsol})  we find that
\begin{equation}
\phi^{(2)}_{\bf k} = {\mu^2\over 56\pi^2} {H_0^2\over k_0^5} a \left( 1 - {k^2 \over 4 k_0^2}\right)
\left( 1 + {k^2\over 3 k_0^2}\right)
\end{equation}
(We have again ignored the decaying mode which arises as a solution to the homogeneous part.)
The corresponding power spectrum for the density field
 $P(k)= |\delta_k|^2 \propto a^2 k^4 |\phi_k|^2 $ will evolve as
\begin{equation}
P^{(2)} (k) \propto (\mu a)^4 q^4 \left( 1 - {1\over 4} q^2\right)^2 \left( 1 + {1\over 3} q^2\right)^2; \quad 
q= {k\over k_0}
\end{equation}
    The power at large spatial scales ($k\to 0$) varies as $k^4$ as discussed before.
    The power has also been generated at smaller scales in the range $k_0<k<2k_0$ with 
    $P^{(2)}(k)$  being a maximum at $k_m\approx 1.54 k_0$ corresponding to the length scale
    $k_m^{-1}\approx 0.65 k_0^{-1}$. 
    Figure \ref{figurepttheory}  shows
    the power spectrum for density field (divided by $a^2$ to eliminate linear growth)
     computed analytically  for a narrow Gaussian
    initial power spectrum centered at $k_0 =1$. The curves are for $(\mu a /56\pi^2)^2 =
    10^{-3}, 10^{-2}, 10^{-1}$ and 1. The similarity between figures \ref{figptsimu}  and
    \ref{figurepttheory}  
    is striking and allows us to understand the simulation results. The key difference
    is that, in the simulations, newly generated power will further produce power at
    $4k_0, 8k_0, ...$ and each of these will give rise to a $k^4$ tail to the right.
    The resultant power will, of course, be more complicated than predicted by our analytic 
    model.   
    The generation of power near this maximum at $k_m^{-1} = 0.65 k_0^{-1}$
     is  visible as a second peak
    in figure  \ref{figurepttheory}  and around 
    $2\pi/k_0\approx  4$ in figure \ref{figptsimu}. 
    
    If we had taken the initial power spectrum to be
    Dirac delta function in the wave vector ${\bf k}$ (rather than on the {\it magnitude} of the wave vector,
    as we have done) the right hand side of Eq.(\ref{approsol}) will contribute
    only when  $({1\over 2}{\bf k}\pm{\bf p}) = {\bf k}_0$. This requires ${\bf p} =0$ and ${\bf k} = 2{\bf k}_0$
    showing that the power is generated exactly at the second harmonic of the wave number.
    Spreading the initial power on a shell of radius $k_0$, spreads the power over different vectors
    leading to the result obtained above. 
    
    Equation (\ref{initcond}) shows that $k_0^3\delta_{\bf k}$ will reach  non linearity for
    $\mu a \approx (3/2)$. The situation is different as regards the gravitational potential
    due to the large numerical factor $56 \pi^2$; the gravitational potential fluctuations are 
    comparable to the original fluctuations only when $\mu a \approx 56 \pi^2$.

    \section{Conclusions}
    
    The purpose of this paper was to draw attention to a possible approach to study nonlinear gravitational clustering. I showed how one can obtain a closed integro-differential equation for the evolution of the gravitational potential in the Fourier space. This equation shows that the nonlinear evolution of the potential at the wave number $\mathbf{k}$ is essentially dictated by a two-mode coupling between modes at
    wave numbers $(1/2)\mathbf{k} + \mathbf{p}$ and  $(1/2)\mathbf{k} - \mathbf{p}$ integrated over all other modes $\mathbf{p}$ with a quadratic kernel. Among other things, this equation is useful in studying how the power injected at a given scale flows to other scales in gravitational clustering. It is seen that the nonlinear evolution leads to a $k^4$ tail at larger spatial scales and pumps power into smaller scales through repeated generation of higher harmonics.
     
     The derivation of the dynamical equation was based on an ansatz which needs to be verified by simulations. We shall report on this aspect as well as on more detailed features of power transfer in a separate publication \cite{stp}. There are several other obvious directions in which this formalism can be developed further. For example, one can incorporate the effect of bound clusters by adding the
     $\phi_{\bf k}\propto k^{-2}$ term due to the monopole moments. One can also improve the accuracy by separating the contribution from different scales in the Fourier space into nonlinear $(k>k_{nl}(t))$ and linear $(k<k_{nl}(t))$ scales (where $k_{nl}(t)$ is the scale that is going nonlinear at time $t$) and dealing with them separately. These issues are under investigation.

 \appendix
   
\section{\label{sphapprox} Spherical approximation}

The purpose of this brief appendix is to show how the formalism developed in the text
connects up with standard spherical approximation used very often in cosmology.
In our language, this approximation consists of assuming that the trajectories are homogeneous; i.e. $ \bld x (t, \bld q) = f (t)\bld q $ where $f(t)$ is to be determined. In this case, the density contrast is
\begin{equation}
\delta_{\bld k} (t)  = \int d^3 \bld q e^{-if(t)\bld k . \bld q} - (2 \pi)^3 \delta_D(\bld k)
 =(2\pi)^3 \delta_D (\bld k) [f^{-3} - 1] \equiv (2 \pi)^3 \delta_D (\bld k)\delta (t) \label{spheapprox}
\end{equation}
where we have defined $\delta(t) \equiv \left[ f^{-3}(t)-1 \right]$ as the amplitude of the density contrast for the $\bld k = 0$ mode. It is now straightforward to compute $A$ and $B$ in Eq.(\ref{exev}).  We have
\begin{equation}
A = 4 \pi G\rho_b \delta^2(t) [(2  \pi)^3 \delta_D(\bld k)] 
\end{equation}
and 
\begin{eqnarray} 
B&=&\int d^3 \bld q (k^aq_a)^2 \dot f^2 e^{-if(k_aq^a)} = -\dot f^2 {\partial^2 \over \partial f^2} [(2 \pi)^3 \delta _D(f \bld k) ] \nonumber \\
 &=& -{4 \over 3} {\dot \delta^2 \over (1 + \delta)} [(2 \pi)^3 \delta_D (\bld k)]
\end{eqnarray}
so that the Eq.~(\ref{exev})  becomes 
\begin{equation}
\ddot\delta + 2 {\dot a \over a} \dot\delta = 4 \pi G \rho_b (1 + \delta) \delta + {4 \over 3} {\dot\delta^2 \over (1 + \delta)} \label{x}
\end{equation}
It is easy to show that this equation is identical to that of standard spherical approximation
governed by $\ddot R = - G M/R^2$ if we take  $(1+ \delta) \propto (a/R)^3$. Thus our formalism allows one to reproduce all the known, standard approximations for nonlinear epochs.


\begin{thebibliography}{99.}
   
   
   \bibitem{tppr}
    T. Padmanabhan:  Physics Reports {\bf 188}, 285 (1990); T.Padmanabhan in  \textit{Dynamics and Thermodynamics of Systems with Long Range Interactions} Eds: T.Dauxois, S.Ruffo, E.Arimondo, M.Wilkens; Lecture Notes in Physics, Springer (2002) [astro-ph/0206131].
   
   
   \bibitem{sample} 
 Chavanis, P-H, in  \textit{Dynamics and Thermodynamics of Systems with Long Range Interactions} Eds: T.Dauxois, S.Ruffo, E.Arimondo, M.Wilkens; Lecture Notes in Physics, Springer (2002); 
E. Follana, V. Laliena, Phys. Rev. {\bf E 61}, 6270 (2000); 
Cooray, A and Sheth, R, Phys.Rept.\textbf{372}:1-129,(2002); [astro-ph/0206508]; 
Scoccimarro, R and Frieman, J, Astrophys.J.\textbf{473}:620,(1996);
H. J. de Vega, N. S'anchez, F. Combes, astro-ph/9807048; 
T.Padmanabhan:  Astrophys. Jour. Supp. , {\bf 71}, 651 (1989);   
Tatekawa, T, astro-ph/0412025;
Buchert, T and Dominguez, A,   Astron.Astrophys.\textbf{438}:443-460 (2005).
     
  
   \bibitem{cosmotext}
   P.J.E. Peebles: \emph{Principles of Physical Cosmology}, (Princeton University Press, New Jersey, 1993);
      T. Padmanabhan: \emph{Structure Formation in the Universe}, (Cambridge University Press, Cambridge 
    1993);
   T. Padmanabhan: \emph{Theoretical Astrophysics, Vol.III: Galaxies and Cosmology}, (Cambridge
   University Press, Cambridge, 2002).
\bibitem{lssu} P.J.E. Peebles: \emph{Large Scale Structure of the Universe}, (Princeton
   University Press, New Jersey, 1980). 
  
      
\bibitem{za} 
Ya.B. Zeldovich, Astron.Astrophys., \textbf{5}, 84,(1970); 
Gurbatov, S. N. et al, MNRAS, \textbf{236}, 385 (1989);  
T.G. Brainerd et al., Astrophys.J., \textbf{418}, 570 (1993);
Matarrese, S et al., MNRAS \textbf{259}:437-452 (1992);
J.S. Bagla, T.Padmanabhan, MNRAS, \textbf{ 266}, 227 (1994) [gr-qc/9304021];
T.Padmanabhan, S.Engineer, Ap. J.,   \textbf{493}, 509 (1998) [astro-ph/9704224];  
S. Engineer  et.al.,  MNRAS  \textbf{314} , 279 (2000) [astro-ph/9812452]; 
for a recent review, see  T.Tatekawa, [astro-ph/0412025].
\bibitem{pi} 
Buchert, T, MNRAS.\textbf{267}:811-820,(1994);
P. Valageas: A  \&A,  {\bf 382}, 477 (2001); A\&A , {\bf 379}, 8 (2001).
\bibitem{nsr}
A.~J.~S. Hamilton et al.,  Ap. J., \textbf{374}, L1 (1991),;  
 T.Padmanabhan et al.,Ap. J.,\textbf{466}, 604 (1996) [astro-ph/9506051];  
  D. Munshi et al., MNRAS,  \textbf{290}, 193 (1997) [astro-ph/9606170];
 J.~S. Bagla, et.al., Ap.J., \textbf{495}, 25 (1998) [astro-ph/9707330];
  N.Kanekar et al., MNRAS, \textbf{ 324}, 988 (2001) [astro-ph/0101562]. 
R. Nityananda, T. Padmanabhan, MNRAS, \textbf{271}, 976 (1994) [gr-qc/9304022]; 
T. Padmanabhan, MNRAS, \textbf{278}, L29 (1996) [astro-ph/9508124]. 

\bibitem{jsbtp1}
   J.S. Bagla and T. Padmanabhan:  Mon Not. R. Astr. Soc., {\bf 286}, 1023 (1997).

 \bibitem{stp} T. Padmanabhan, Suryadeep Ray, astro-ph/0511596.
 
 \bibitem{pines}
   D.Pines and D.Bohm, Phys.Rev. {\bf 85}, 338 (1952).
   
 \bibitem{tpprob}
   T. Padmanabhan: \emph{Cosmology 
     and Astrophysics through Problems}, (Cambridge University Press, Cambridge 1996).
     
 
   
 
   
   
 
 
 \end{thebibliography}
 \end{document}